\batchmode
\makeatletter
\def\input@path{{D:/rsegevOnD/CURRENTonDESK/ConferencesAndVisits/ASME06/ArXiv//}}
\makeatother
\documentclass[11pt,english,reqno]{amsart}
\usepackage[T1]{fontenc}
\usepackage{amssymb}

\makeatletter
 \theoremstyle{plain}
 
 \numberwithin{equation}{section} 
 \numberwithin{figure}{section} 

\usepackage{graphicx}
\usepackage{graphpap}

\usepackage{babel}
\makeatother
\begin{document}
\newcommand{\reals}{\mathbb{R}}
\newcommand{\rthree}{\reals^{3}}
\newcommand{\fall}{,\quad\text{for all}\quad}
\newcommand{\les}{\leqslant}
\newcommand{\ges}{\geqslant}
\newcommand{\dee}{\mathrm{\mathrm{d}}}
\newcommand{\from}{\colon}
\newcommand{\tto}{\longrightarrow}
\newcommand{\abs}[1]{\left|#1\right|}
\newcommand{\isom}{\cong}
\newcommand{\comp}{\circ}
\newcommand{\cl}[1]{\overline{#1}}
\newcommand{\fun}{\varphi}
\newcommand{\interior}{\mathrm{Int\,}}
\newcommand{\sign}{\mathrm{sign\,}}
\newcommand{\dimension}{\mathrm{dim\,}}
\newcommand{\esssup}{\mathrm{ess}\,\sup}
\newcommand{\ess}{\mathrm{{ess}}}
\newcommand{\kernel}{\mathrm{Kernel\,}}
\newcommand{\support}{\mathrm{Supp}\,}
\newcommand{\image}{\mathrm{Image\,}}
\newcommand{\incl}{\iota}
\newcommand{\rest}{\incl^{*}}
\newcommand{\proj}{\pi}
\newcommand{\ino}[1]{\int\limits _{#1}}
\newcommand{\half}{\frac{1}{2}}
\newcommand{\shalf}{{\scriptstyle \half}}
\newcommand{\empt}{\varnothing}
\newcommand{\paren}[1]{\left(#1\right)}
\newcommand{\bigp}[1]{\bigl(#1\bigr)}
\newcommand{\braces}[1]{\left\{  #1\right\}  }
\newcommand{\sqbr}[1]{\left[#1\right]}
\newcommand{\norm}[1]{\|#1\|}
\newcommand{\dual}{^{*}}
\newcommand{\trps}{^{T}}
\newcommand{\wh}[1]{\widehat{#1}}
\newcommand{\pis}{x}
\newcommand{\pib}{X}
\newcommand{\body}{B}
\newcommand{\bdry}{\partial}
\newcommand{\gO}{\varOmega}
\newcommand{\reg}{\gO}
\newcommand{\bdom}{\bdry\gO}
\newcommand{\bndo}{\partial\gO}
\newcommand{\cloo}{\cl{\gO}}
\newcommand{\nor}{\nu}
\newcommand{\dA}{\,\dee A}
\newcommand{\dV}{\,\dee V}
\newcommand{\eps}{\varepsilon}
\newcommand{\vect}{v}
\newcommand{\vs}{\mathcal{W}}
\newcommand{\vbase}{e}
\newcommand{\vf}{w}
\newcommand{\avf}{u}
\newcommand{\stn}{\varepsilon}
\newcommand{\rig}{r}
\newcommand{\rigs}{\mathcal{{R}}}
\newcommand{\qrigs}{\!/\!\rigs}
\newcommand{\dis}{\chi}
\newcommand{\fc}{F}
\newcommand{\st}{\sigma}
\newcommand{\bfc}{b}
\newcommand{\sfc}{t}
\newcommand{\stm}{S}
\newcommand{\sts}{\varSigma}
\newcommand{\ebdfc}{T}
\newcommand{\optimum}{S}
\newcommand{\opt}{\mathrm{opt}}
\newcommand{\cee}[1]{C^{#1}}
\newcommand{\lone}{L^{1}}
\newcommand{\linf}{L^{\infty}}
\newcommand{\ofbdo}{(\bndo)}
\newcommand{\ofclo}{(\cloo)}
\newcommand{\vono}{(\gO,\rthree)}
\newcommand{\vonbdo}{(\bndo,\rthree)}
\newcommand{\vonclo}{(\cl{\gO},\rthree)}
\newcommand{\strono}{(\gO,\reals^{6})}
\newcommand{\sob}{W_{1}^{1}}
\newcommand{\sobb}{\sob(\gO,\rthree)}
\newcommand{\lob}{\lone(\gO,\rthree)}
\newcommand{\lib}{\linf(\gO,\reals^{12})}
\newcommand{\ofO}{(\gO)}
\newcommand{\oneo}{{1,\gO}}
\newcommand{\onebdo}{{1,\bndo}}
\newcommand{\info}{{\infty,\gO}}
\newcommand{\infclo}{{\infty,\cloo}}
\newcommand{\infbdo}{{\infty,\bndo}}
\newcommand{\ld}{}
\newcommand{\ldo}{\ld\ofO}
\newcommand{\trace}{\gamma}
\newcommand{\pr}{\proj_{\rigs}}
\newcommand{\pq}{\proj}
\newcommand{\qr}{\,/\,\reals}
\newcommand{\ssx}{S}
\newcommand{\smap}{s}
\newcommand{\smat}{\chi}
\newcommand{\sx}{e}
\newcommand{\snode}{P}
\newcommand{\node}{p}
\newcommand{\elem}{e}
\newcommand{\nel}{L}
\newcommand{\el}{l}
\newcommand{\ipln}{\phi}
\newcommand{\ndof}{D}
\newcommand{\dof}{d}
\newcommand{\nldof}{N}
\newcommand{\ldof}{n}
\newcommand{\lvf}{\chi}
\newcommand{\lfc}{\varphi}
\newcommand{\amat}{A}
\newcommand{\snomat}{E}
\newcommand{\femat}{E}
\newcommand{\tmat}{T}
\newcommand{\fvec}{f}
\newcommand{\snsp}{\mathcal{S}}
\newcommand{\slnsp}{\Phi}
\newcommand{\ro}{r_{1}}
\newcommand{\rtwo}{r_{2}}
\newcommand{\rth}{r_{3}}
\newcommand{\aro}{S_{1}}
\newcommand{\art}{S_{2}}
\newcommand{\mo}{m_{1}}
\newcommand{\mt}{m_{2}}
\title{Optimal Stresses in Structures}

\author{Reuven Segev and Gal deBotton}

\address{Pearlstone Center for Aeronautical Engineering Studies, \\
Department of Mechanical Engineering, \\
Ben-Gurion University, Beer-Sheva, Israel, \\
rsegev@bgu.ac.il}

\dedicatory{In memory of Israel Gilad (1949 -- 2005)}

\date{\today}


\begin{abstract}
For a given external loading on a structure we consider the optimal
stresses. Ignoring the material properties the structure may have,
we look for the distribution of internal forces or stresses that is
in equilibrium with the external loading and whose maximal component
is the least. We present an expression for this optimal value in terms
of the external loading and the matrix relating the external degrees
of freedom and the internal degrees of freedom. The implementation
to finite element models consisting of elements of uniform stress
distributions is presented. Finally, we give an example of stress
optimization for of a two-element model of a cylinder under external
traction.
\end{abstract}
\maketitle

\section{INTRODUCTION}

This paper presents an analysis of optimal stresses in structures
under given loadings. In \cite{S03,S04,S05}, optimal stress distributions
for continuous bodies were considered. Although the problem for a
continuous body is more difficult mathematically, the corresponding
analysis for a structure having a finite number of degrees of freedom
is more relevant for engineering applications.

From the point of view of statics, engineering structures---starting
from simple trusses all the way to finite element models used for
stress analysis of continuous bodies, and evidently, continuous models
of bodies---are predominantly statically indeterminate. Mathematically,
this means that we have more unknown parameters describing the stress
distribution in the structure under consideration than equilibrium
equations. This mathematical problem is solved usually by the introduction
of constitutive relations and by coupling the statics problem with
kinematics.

The work presented here takes a different approach to statically indeterminate
problems. Remaining within the framework of statics, we do not specify
any constitutive relations and look for the values of the unknown
components that satisfy the equilibrium conditions and for which the
maximal component is the least. Specifically, the problem may be stated
as follows. Let $\fvec_{\dof}$, $\dof=1,\dots,\ndof$, be the components
of the known external loading vector on the structure where $\ndof$
is the number of degrees of freedom the structure has, and let $\lfc_{\ldof}$,
$\ldof=1,\dots,\nldof$, be the components of the unknown vector of
internal forces---stress-like entities. As we consider statically
indeterminate problems, $\nldof>\ndof$. The equations of equilibrium
will be of the form\begin{equation}
\amat\trps(\lfc)=\fvec,\quad\text{or}\quad\amat\trps_{\dof\ldof}\lfc_{\ldof}=\fvec_{\dof},\end{equation}
where we use the summation convention and the reason we write the
matrix as $\amat\trps$ rather than simply $\amat$ will be made clear
below. Thus, letting $\lfc_{\max}=\max_{\ldof}\abs{\lfc_{\ldof}}$
, we are looking for \begin{equation}
\optimum_{\fvec}^{\opt}=\min_{\lfc}\braces{\lfc_{\max}},\end{equation}
where the minimum is taken over all $\lfc$ satisfying $\amat\trps(\lfc)=\fvec$.

Our basic result states that \begin{equation}
\optimum_{\fvec}^{\opt}=\max_{\vf}\frac{\abs{\fvec_{\dof}\vf_{\dof}}}{\sum_{\ldof}\abs{\amat_{\ldof\dof}\vf_{\dof}}}\,,\end{equation}
where the maximum is taken over all global virtual displacement vectors
$\vf=(\vf_{1},\dots,\vf_{\ndof})$ of the structure, so $\fvec_{\dof}\vf_{\dof}$
is the virtual work performed by the external force vector.

A related quantity that we consider is the \emph{stress sensitivity}
of the structure defined as follows. Assuming that the internal forces
have the same physical dimension as the external forces (dimensions
of forces or forces divided by area), consider the ratio\begin{equation}
K_{\fvec}=\frac{\optimum_{\fvec}^{\opt}}{\max_{\dof}\abs{\fvec_{\dof}}}\,.\end{equation}
Thus, $K_{\fvec}$ measures the sensitivity of the structure to the
external force $\fvec$. Next, we let the external force vary and
we look for the worst possible ratio. The stress sensitivity of the
structure is defined as\begin{equation}
K=\max_{\fvec}K_{\fvec}=\max_{\fvec}\braces{\frac{\optimum_{\fvec}^{\opt}}{\max_{\dof}\abs{\fvec_{\dof}}}}.\label{eq:scf}\end{equation}
It is shown in Section~\ref{sub:The-Stress-Sensitivity} that \begin{equation}
K=\max_{\vf}\frac{\sum_{\dof}\abs{\vf_{\dof}}}{\sum_{\ldof}\abs{\amat_{\ldof\dof}\vf_{\dof}}}\,.\end{equation}
We emphasize that $K$ is a geometric, kinematic property of the structure,
i.e.., independent of material properties, loading conditions, etc.

The paper is outlined as follows. We start with the notation and basic
facts regarding statically indeterminate structures. Then, we prove
the results stated above. For the internal force vector $\lfc$, the
value $\lfc_{\max}$ is represented as a norm, specifically, the dual
of the norm $\norm{\lvf}=\sum_{\ldof}\abs{\lvf_{\ldof}}$ that we
use for internal displacements $\lvf$. The basic tool we use is the
norm preserving extension of functionals (the simplified, finite dimensional
case of the Hahn-Banach theorem). Next, we present some details regarding
the application of the method to the case of finite element models
consisting of elements having uniform stress distributions. Finally,
in a way of example, we consider the case of a two elements model
of a thick cylinder under external symmetric loading.

\section{STATICALLY INDETERMINATE STRUCTURES}

An elementary example for the type of structures we consider is shown
in Fig.~1. Our method applies to a lot more complicated structures
including a large variety of finite element models.

\begin{figure}

\begin{center}

\hspace*{4mm}\begin{picture}(250,250)


\put(0,0){\includegraphics[scale=0.7]{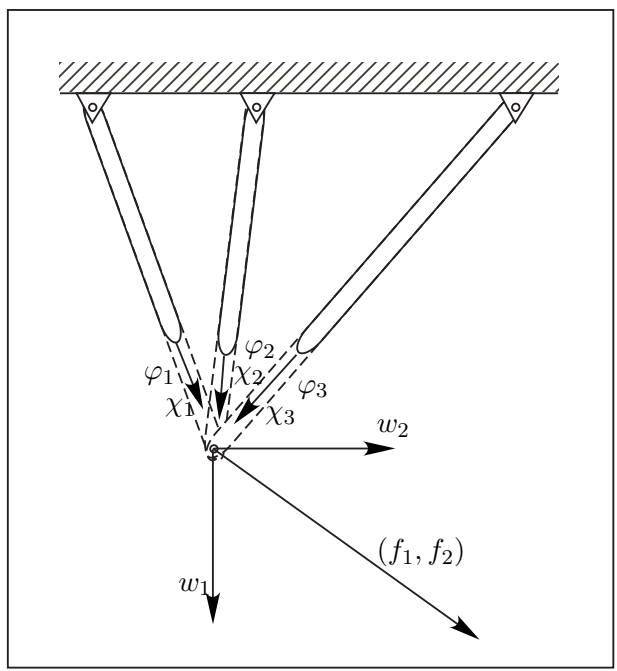}}











\put(140,42){$(\fvec_1,\fvec_2)$}

\put(140,90){$\vf_2$}

\put(65,30){$\vf_1$}

\put(52,110){$\lfc_1$}

\put(60,98){$\lvf_1$}

\put(90,120){$\lfc_2$}

\put(86,110){$\lvf_2$}

\put(110,105){\smash{$\lfc_3$}}

\put(98,95){$\lvf_3$}



\end{picture}

\caption{A SIMPLE INDETERMINATE STRUCTURE}

\end{center}

\end{figure}

\subsection{Kinematics}

The structure is assumed to have $\ndof$ degrees of freedom. This
means that we have a $\ndof$-dimensional vector space $\vs$ containing
the \emph{external} infinitesimal virtual displacements (generalized
velocities) that are compatible with the displacements boundary conditions
at the supports and the various constraints implied by the structural
connections. A generic virtual displacement in $\vs$ will be denoted
by $\vf$. Thus in our model example of Fig.~1, the structure has
two degrees of freedom and a generic virtual displacement is of the
form $\vf=(\vf_{1},\vf_{2}).$ Each external degree of freedom induces
a base vector in $\vs$.

Next we consider the space $\snsp$ containing noncompatible, or \emph{internal},
infinitesimal virtual displacements of the structure. Such an internal
deformation field of our model example is shown in Fig.~1. Here,
the constraints of the structural interconnections of the various
structural elements are not kept (the connection at the bottom joint
in the figure). In particular, constant strains within structural
elements may be represented as internal virtual displacement fields.
A generic internal virtual displacement field will be denoted as $\lvf$.
Clearly, internal deformations have more degrees of freedom than the
external ones. We assume formally that $\nldof$, the dimension of
$\snsp$, is strictly larger than $\ndof$. The various internal degrees
of freedom induce base vectors in $\snsp$. (Returning to our model
example, we note that the internal degrees of freedom, i.e, the base
vectors in $\snsp$, may be unit changes in lengths of the bars or
unit axial strains in the bars. It will be convenient also to define
for a uniform strain structural member, such as a bar in the example,
base vectors in $\snsp$ consisting of unit strain components in the
element multiplied by its volume as in Section \ref{sec:Uniform-Stress-Finite-Elements}.)

An important role in the kinematics of the structure is played by
the \emph{interpolation mapping} \begin{equation}
\amat\from\vs\tto\snsp.\end{equation}
This mapping associates an internal deformation vector with every
external deformation. Clearly, as $\nldof>\ndof$, not all internal
deformations may be obtained as images of external deformations under
$\amat$. We further note that in case the structure is not supported
distinct displacements fields that differ by a rigid displacement
field induce the same strain field. While unsupported bodies may be
considered following the methods of \cite{S04}, we simplify the analysis
and assume that the supports prevent such rigid displacement fields.
Thus, we assume mathematically that the interpolation mapping is one-to-one,
so the matrix of $\amat$ is of full rank $\ndof$.

\subsection{Statics}

An external force $\fvec$ performs virtual work (power) for various
external displacements. Denoting by $\fvec_{d}$ the component of
the force dual to the degree of freedom $\vf_{\dof}$, the virtual
work may be written as \begin{equation}
\fvec(\vf)=\fvec_{\dof}\vf_{d}.\end{equation}
In other words, we regard an external force as a linear functional
\begin{equation}
\fvec\from\vs\tto\reals\end{equation}
and the collection of all external forces is the dual space $\vs\dual$
containing all real valued linear mappings defined on $\vs$.

In analogy, an internal force $\lfc$ performs virtual work for virtual
internal deformation fields. Denoting the component of the internal
force $\lfc$ corresponding to the component $\lvf_{\ldof}$, by $\lfc_{\ldof}$,
$\ldof=1,\dots,\nldof$, we write for the internal virtual work performed
by $\lfc$ for the virtual displacement $\lvf$\begin{equation}
\lfc(\lvf)=\lfc_{\ldof}\lvf_{n}.\end{equation}
 Thus, an internal force is a linear mapping\begin{equation}
\lfc\from\snsp\tto\reals,\end{equation}
i.e., $\lfc$ belongs to the space $\snsp\dual$ of real valued linear
mapping on $\snsp$. Note that in case the component $\lvf_{n}$ indicates
a constant component of the strain in some structural element multiplied
by its volume, then $\lfc_{\ldof}$ indicates the corresponding stress
component (see Section \ref{sec:Uniform-Stress-Finite-Elements}).

The principle of virtual work serves as the condition for equilibrium
within the framework of the structural model. Using the notation introduced
above it states that\begin{equation}
\lfc(\amat(\vf))=\fvec(\vf),\end{equation}
for all external vector fields $\vf$ in $\vs$. Using matrix notation
where we keep the same symbol for a linear mapping (or a vector) and
its corresponding matrix (or the corresponding column vector) the
principle of virtual work is written as\begin{equation}
\lfc\trps\amat\vf=\fvec\trps\vf.\end{equation}
 Thus, the principle of virtual work, or equivalently the equilibrium
condition, may be written in any of the following forms\begin{equation}
\lfc\comp\amat=\fvec,\quad\amat\trps\lfc=\fvec,\quad\amat_{\ldof\dof}\lfc_{\ldof}=\fvec_{\dof},\quad\amat\dual(\lfc)=\fvec,\label{eq:equil}\end{equation}
where in the last equation above we used the dual mapping $\amat\dual\from\snsp\dual\tto\vs\dual$
defined by the condition $\amat\dual(\lfc)(\vf)=\lfc(\amat(\vf))$
and whose matrix is the transpose of that of $\amat$ as expected.

Given an external force $\fvec$, the equilibrium conditions (\ref{eq:equil})
provide a system of $\ndof$ equations for the $\nldof$ components
of the internal force $\lfc$. As it was assumed that $\nldof$ is
strictly larger than $\ndof$ and that $\amat$ is one-to-one, this
system of equations cannot determine $\lfc$ uniquely. In fact, there
is an $(\nldof-\ndof)$-dimensional vector space $\slnsp=(\amat\dual)^{-1}\braces{\fvec}$
of solutions to the equilibrium problem.

\section{OPTIMAL SOLUTIONS}

For given structures, where the material properties of the various
structural elements are known, the constitutive relations provide
the additional information so the internal force vector can be calculated
uniquely for any given external force $\fvec$. Here however, we consider
the situation where no constitutive relations are given a-priori,
and among all solutions $\lfc$ in $\slnsp$, we look for the least
bound on the maximal component.

Specifically, for each internal force $\lfc$, we set\begin{equation}
\norm{\lfc}^{\infty}=\lfc_{\max}=\max_{\ldof}\abs{\lfc_{\ldof}},\end{equation}
and we look for\begin{equation}
\optimum_{\fvec}^{\opt}=\min_{\lfc}\braces{\norm{\lfc}^{\infty}}=\min_{\lfc}\braces{\max_{\ldof}\abs{\lfc_{\ldof}}},\end{equation}
where the minimum is taken over all internal forces $\lfc$ satisfying
$\amat\dual(\lfc)=\fvec$. Thus, for the case were the components
of the internal force represent stresses in the various structural
elements, we are looking for the least bound on the discretized approximating
stress field.

In order to evaluate $\optimum_{\fvec}^{\opt}$ directly on the basis
of its definition, one would have to generate the space of solutions
$\slnsp$ and then evaluate the optimal bound in that space. The analysis
we present in the sequel will give an expression for $\optimum_{\fvec}^{\opt}$
that does not require the solution of the equilibrium equations (\ref{eq:equil}).

\subsection{A Solution as an Extension of a Functional}

Since the interpolation mapping $\amat$ is one-to-one, its inverse\begin{equation}
\amat^{-1}\from\image\amat\tto\vs\end{equation}
is well defined on its image, a subspace of $\snsp$. Given an external
force $\fvec$, we consider \begin{equation}
\wh{\fvec}=\fvec\comp\amat^{-1}\from\image\amat\tto\reals,\end{equation}
a linear mapping defined on the subspace $\image\amat$. Note that
the equilibrium condition $\fvec(\vf)=\lfc(\amat(\vf))$, for a linear
functional $\lfc$ on $\snsp$ may be written as\begin{equation}
\lfc(\lvf)=\fvec\comp\amat^{-1}(\lvf)=\wh{\fvec}(\lvf)\label{eq:ExtMeansPVW}\end{equation}
for all compatible internal fields $\lvf$ in $\image\amat$. In other
words, a solution of the equilibrium equations is a linear mapping
$\lfc$ defined on $\snsp$ that agrees with $\wh{\fvec}=\fvec\comp\amat^{-1}$
on the subspace $\image\amat$. Thus, $\lfc$ is an \emph{extension}
of $\wh{\fvec}$ to the entire space $\snsp$.

It is noted that algebraically, given the linear functional $\wh{\fvec}$
on the subspace, it is straightforward to generate an extension of
it to the entire space. In fact, it is sufficient to show this for
the case where $\nldof-\ndof=1$, so we have to extend $\wh{\fvec}$
to a space having one more dimension. In the general case where $\nldof-\ndof$
is any other finite number, the procedure can be carried out inductively
adding one dimension at a time. To generate such an extension $\lfc$,
one can choose an internal displacement vector $\lvf_{1}$ that does
not belong to $\image\amat$ and give an arbitrary value to $c=\lfc(\lvf_{1})$.
Any vector $\lvf$ in the larger space may be written as a linear
combination\begin{equation}
\lvf=\lvf_{0}+a\lvf_{1}\end{equation}
for a compatible internal virtual displacement $\lvf_{0}$ in $\image\amat$
and a real number $a$. Thus, for any extension $\lfc$ we have\begin{align}
\lfc(\lvf) & =\lfc(\lvf_{0}+a\lvf_{1})\nonumber \\
 & =\lfc(\lvf_{0})+a\lfc(\lvf_{1})\nonumber \\
 & =\wh{\fvec}(\lvf_{0})+ac.\label{eq:algebraicExtension}\end{align}

\subsection{Norms}

Recalling that we are looking for a solution of the equilibrium equations
that minimizes $\norm{\lfc}^{\infty}=\max_{\ldof}\abs{\lfc_{\ldof}}$,
we mention a number of useful properties of this norm on the space
of internal forces. Consider the norm on the space $\snsp$ given
by\begin{equation}
\norm{\lvf}_{1}=\sum_{\ldof}\abs{\lvf_{\ldof}}.\label{eq:L1-Norm}\end{equation}
Then, the following holds\begin{equation}
\norm{\lfc}^{\infty}=\max_{\lvf}\frac{\abs{\lfc(\lvf)}}{\norm{\lvf}_{1}}\,.\end{equation}
This relation between the norm for internal forces and the norm for
internal virtual displacements is all we need. In fact, the following
analysis applies to other criteria for optimization of the internal
forces, i.e., criteria given by other norms say $\norm{\lfc}$. To
do this, one should determine the norm on the space of internal displacements
to which $\norm{\lfc}$ is dual, i.e., determine the norm $\norm{\lvf}$
such that\begin{equation}
\norm{\lfc}=\max_{\lvf}\frac{\abs{\lfc(\lvf)}}{\norm{\lvf}}\,.\label{eq:DualNorm}\end{equation}
Thus, in the sequel we use a generic norm $\norm{\lvf}$ for internal
displacements and the corresponding dual norm $\norm{\lfc}$ for internal
forces satisfying Equation (\ref{eq:DualNorm}). In fact, once the
norm $\norm{\lvf}$ is established as the one for which our optimality
condition is a dual norm, the norm on the space of internal displacement
fields (rather than the one on the space of internal forces) plays
the central role as will be seen below. It is noted that if the optimality
criterion is given in terms of a norm $\norm{\lfc}$, the associated
norm on the space of internal displacements can be found by (see \cite[p. 186]{Taylor})\begin{equation}
\norm{\lvf}=\max_{\lfc}\frac{\abs{\lfc(\lvf)}}{\norm{\lfc}}\,.\label{eq:normDblDual}\end{equation}

For external virtual displacements we may also consider the norm\begin{equation}
\norm{\vf}_{1}=\sum_{\dof}\abs{\vf_{\dof}},\end{equation}
and use the dual norm\begin{equation}
\norm{\fvec}^{\infty}=\max_{\vf}\frac{\abs{\fvec(\vf)}}{\norm{\vf}_{1}}=\max_{\dof}\abs{\fvec_{\dof}}\end{equation}
for external force vectors.

Contrary to the previous paragraph, we will find it useful in the
analysis of optimal internal forces below to use for external virtual
displacements the norm\begin{equation}
\norm{\vf}=\norm{\amat(\vf)},\label{eq:NormOnW}\end{equation}
where on the right we use the norm on $\snsp$. The corresponding
dual norm for external forces is therefore\begin{equation}
\norm{\fvec}=\max_{\vf}\frac{\abs{\fvec(\vf)}}{\norm{\vf}}=\max_{\vf}\frac{\abs{\fvec(\vf)}}{\norm{\amat(\vf)}}\,.\end{equation}

\subsection{Optimal Extensions}

Considering the linear functional $\wh{\fvec}=\fvec\comp\amat^{-1}\from\image\amat\to\reals$,
we may evaluate its dual norm relative to the norm (\ref{eq:L1-Norm})
on $\image\amat$. Thus, \begin{equation}
\norm{\wh{\fvec}}^{\infty}=\max_{\lvf\in\image\amat}\frac{\abs{\wh{\fvec}(\lvf)}}{\norm{\lvf}_{1}}\end{equation}
where it is noted that the maximum is evaluated for all $\lvf$ in
$\image\amat$ (and not the entire space $\snsp$). The Hahn-Banach
theorem of functional analysis states that there is a linear functional
$\lfc_{\fvec}^{\opt}\from\snsp\tto\reals$ (i.e., defined on the space
$\snsp$) such that \begin{equation}
\lfc_{\fvec}^{\opt}(\lvf)=\wh{\fvec}(\lvf)\end{equation}
for all $\lvf$ in $\image\amat$, and \begin{equation}
\norm{\lfc_{\fvec}^{\opt}}^{\infty}=\max_{\lvf\in\snsp}\frac{\abs{\lfc_{\fvec}^{\opt}(\lvf)}}{\norm{\lvf}_{1}}=\norm{\wh{\fvec}}^{\infty}=\max_{\lvf\in\image\amat}\frac{\abs{\wh{\fvec}(\lvf)}}{\norm{\lvf}_{1}}\,.\label{eq:HB-EqualNorms}\end{equation}
In other words, $\lfc$ extends $\wh{\fvec}$ without increasing its
norm. It was mentioned earlier that extending the functional one dimension
at a time is simple (see Equation (\ref{eq:algebraicExtension}).
Appendix~A presents the construction for the addition of one dimension
to the domain without increasing the norm of the linear functional.
In fact, the Hahn-Banach theorem asserts that this can be done for
infinite dimensional spaces also and is used in the continuum counterpart
of this analysis in \cite{S03,S04,S05}.

We end this subsection by noting that \begin{equation}
\norm{\wh{\fvec}}^{\infty}=\max_{\lvf\in\image\amat}\frac{\abs{\wh{\fvec}(\lvf)}}{\norm{\lvf}_{1}}=\max_{\vf\in\vs}\frac{\abs{\fvec(\vf)}}{\norm{\amat(\vf)}}=\norm{\fvec}.\label{eq:normFHatisNormF}\end{equation}

\subsection{The Equation for the Optimum}

Returning to the expression for the optimum, we note that in general
for any internal force $\lfc$ that extends $\wh{\fvec}$\begin{align}
\norm{\wh{\fvec}}^{\infty} & =\max_{\lvf\in\image\amat}\frac{\abs{\wh{\fvec}(\lvf)}}{\norm{\lvf}}\\
 & =\max_{\lvf\in\image\amat}\frac{\abs{\lfc(\lvf)}}{\norm{\lvf}}\quad\text{(by the principle of virtual work)}\\
 & \les\max_{\lvf\in\snsp}\frac{\abs{\lfc(\lvf)}}{\norm{\lvf}}=\norm{\lfc}^{\infty}.\end{align}
Thus, since for $\lfc_{\fvec}^{\opt}$, $\norm{\lfc_{\fvec}^{\opt}}^{\infty}=\norm{\wh{\fvec}}^{\infty}=\norm{\fvec},$
as in Eqs.~(\ref{eq:HB-EqualNorms}) and (\ref{eq:normFHatisNormF}),
we have\begin{equation}
\norm{\lfc_{\fvec}^{\opt}}^{\infty}=\norm{\fvec}=\min_{\lfc}\norm{\lfc}^{\infty},\end{equation}
where the minimum is taken over all extensions $\lfc$ of $\wh{\fvec}$.
Since an extension $\lfc$ of $\wh{\fvec}$ satisfies the equilibrium
condition as in Eq. (\ref{eq:ExtMeansPVW}), we conclude that\begin{equation}
\min_{\fvec=\amat\dual(\lfc)}\braces{\max_{\ldof}\abs{(\lfc_{\fvec}^{\opt})_{\ldof}}}=\norm{\lfc_{\fvec}^{\opt}}^{\infty}=\norm{\fvec}.\end{equation}
It follows that\begin{equation}
\optimum_{\fvec}^{\opt}=\max_{\vf\in\vs}\frac{\abs{\fvec(\vf)}}{\norm{\amat(\vf)}}=\max_{\vf\in\vs}\frac{\abs{\fvec_{d}\vf_{d}}}{\sum_{\ldof}\abs{\amat_{\ldof\dof}\vf_{\dof}}}\,.\label{eq:ForOptimum}\end{equation}

\subsection{The Stress Sensitivity\label{sub:The-Stress-Sensitivity}}

We now derive the expression for the stress sensitivity of the structure.
Recalling its definition in Eq. (\ref{eq:scf}) \[
K=\max_{\fvec}\braces{\frac{\optimum_{\fvec}^{\opt}}{\max_{\dof}\abs{\fvec_{\dof}}}},\]
we substitute the expression for the optimum internal force to get\begin{align}
K & =\max_{\fvec}\braces{\frac{1}{\max_{\dof}\abs{\fvec_{\dof}}}\braces{\max_{\vf\in\vs}\frac{\abs{\fvec_{d}\vf_{d}}}{\sum_{\ldof}\abs{\amat_{\ldof\dof}\vf_{\dof}}}}}\\
 & =\max_{\vf\in\vs}\braces{\frac{1}{\sum_{\ldof}\abs{\amat_{\ldof\dof}\vf_{\dof}}}\braces{\max_{\fvec}\frac{\abs{\fvec_{d}\vf_{d}}}{\max_{\dof}\abs{\fvec_{\dof}}}}}.\end{align}
 We may use now Eq. (\ref{eq:normDblDual}) and arrive at\begin{equation}
K=\max_{\vf\in\vs}\frac{\sum_{\dof}\abs{\vf_{\dof}}}{\sum_{\ldof}\abs{\amat_{\ldof\dof}\vf_{\dof}}}\,.\end{equation}

\section{APPLICATION TO FINITE ELEMENTS\label{sec:Uniform-Stress-Finite-Elements}}

A typical situation where one would like to apply the foregoing analysis
is a finite element model of a continuous body. Thus, we briefly describe
here some additional details for the simple situation of a finite
element model where it is assumed that the stress is  uniform within
each element. We do not consider here the question of approximation
of the solution to the continuum problem by finite elements and take
the finite element model as a given structure.

Let $\nel$ be the number of elements, $\elem_{\el}$ the $\el$-th
element, and $\st_{ij}^{\el}$ the components of the uniform stress
in that element. We want the collection of $\st_{ij}^{\el}$ for the
various elements $\el$ and various components $i,j$ to be the components
of our internal force vector. Thus, the index $\ldof$ is replaced
by the collection of 3 indices $\el,i,j$, $\lfc_{ij}^{\el}=\st_{ij}^{\el}$,
and\begin{equation}
\norm{\lfc}=\max_{i,j,\el}\abs{\st_{ij}^{\el}}\end{equation}
is the quantity that we want to minimize.

The internal degrees of freedom $\lvf_{ij}^{\el}$ should be chosen
such that $\sum_{\el,i,j}\st_{ij}^{\el}\lvf_{ij}^{\el}$ is the virtual
work of the internal force for the given internal displacements. Writing\begin{equation}
U=\ino{\cup_{\el}\elem_{\el}}\st_{ij}\eps_{ij}\dV=\sum_{\el}\st_{ij}^{\el}\ino{\elem_{\el}}\eps_{ij}\dV,\end{equation}
where $\eps_{ij}$ denotes the linear strain field, one realizes that
the internal degrees of freedom are given by\begin{equation}
\lvf_{ij}^{\el}=\ino{\elem_{\el}}\eps_{ij}\dV.\end{equation}
For uniform strain elements $\lvf_{ij}^{\el}=V_{\el}\eps_{ij}^{\el}$
(no sum on $\el$), where $\eps_{ij}^{\el}$ are the components of
the uniform strain in the element $\elem_{\el}$ and $V_{\el}$ is
its volume. The norm $\norm{\lvf}$ is therefore\begin{equation}
\norm{\lvf}=\sum_{i,j,\el}\abs{\ino{\elem_{\el}}\eps_{ij}\dV}\end{equation}
and for uniform strain elements\begin{equation}
\norm{\lvf}=\sum_{i,j,\el}\abs{\eps_{ij}^{\el}}V_{\el}.\end{equation}

It is a standard procedure in the construction of finite element models
to use an array $\femat_{ij\dof}^{\el}$ such that\begin{equation}
\lvf_{ij}^{\el}=\ino{\elem_{\el}}\eps_{ij}\dV=\femat_{ij\dof}^{\el}\vf_{\dof},\end{equation}
where the array $\femat_{ij\dof}^{\el}$ is generated using the relation
between the external degrees of freedom and the displacements at the
nodes of the the element $\elem_{\el}$, using the shape functions
to obtain the displacement field within the element, and using the
discrete forms of the differentiation and integration operators in
order to obtain the strain components and their integrals. Thus, the
array $\femat_{ij\dof}^{\el}$ replaces the matrix $\amat_{\ldof\dof}$
in our expression for the optimum. We conclude that\begin{equation}
\optimum_{\fvec}^{\opt}=\norm{\lfc_{\fvec}^{\opt}}=\min_{\st}\braces{\max_{\el,i,j}\abs{\st_{ij}^{\el}}}=\max_{\vf\in\vs}\frac{\abs{\fvec_{d}\vf_{d}}}{\sum_{\el,i,j}\abs{\femat_{ij\dof}^{\el}\vf_{\dof}}}\,.\end{equation}

\section{EXAMPLE}

We consider the following 2-dimensional example. The structure is
a finite element model of a cylinder of inner radius $\ro$ and outer
radius $\rth$ under external normal traction $p$ at the outer boundary
(see Fig. 2).

\begin{figure}

\begin{center}

\hspace*{4mm}\begin{picture}(225,230)


\put(1,-1){\includegraphics[scale=0.4]{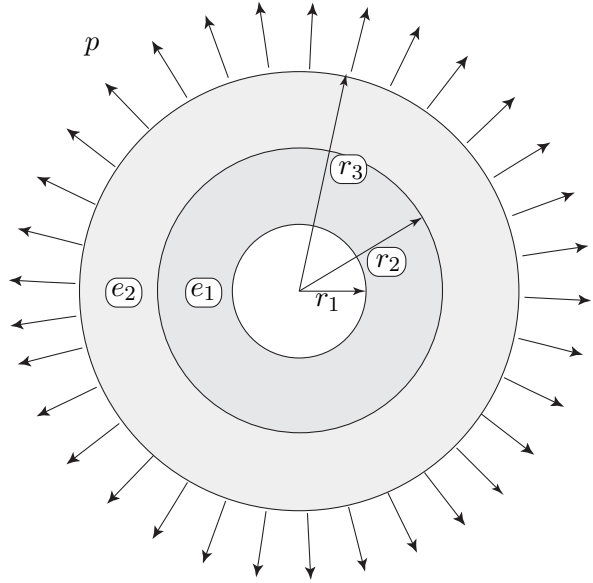}}





\put(126,153){$\rth$}

\put(30,200){$p$}

\put(70,107){$\elem_1$}

\put(140,118){$\rtwo$}

\put(40,107){$\elem_2$}

\put(117,103){{$\ro$}}

\end{picture}

\caption{A 2 ELEMENT MODEL OF A CYLINDER}

\end{center}

\end{figure}

The finite element model consists of two uniform stress elements
$\elem_{1}$ and $\elem_{2}$, corresponding to the regions $\ro\les r\les\rtwo$,
and $\rtwo\les r\les\rth$, respectively, where $r_{2}=(\ro+\rth)/2$.
Due to the cylindrical symmetry the problem has three degrees of freedom
and a typical external virtual displacement is of the form $\vf=(\vf_{1},\vf_{2},\vf_{3})$,
where $\vf_{\dof}$ is the radial displacement at $r_{\dof}$. The
corresponding components of an external force $\fvec$ are given as\begin{equation}
\fvec_{\dof}=2\pi r_{\dof}p_{\dof},\quad\text{(no summation),}\quad\dof=1,2,3,\end{equation}
where, $p_{d}$ is the applied load at $r_{d}$. Thus, for the case
under consideration $\fvec=(0,0,2\pi\rth p)$. The space of internal
virtual displacements $\snsp$ will be 4-dimensional and an internal
displacement will be of the form $\lvf_{1}=\aro(\eps_{r})_{1}$, $\lvf_{2}=\aro(\eps_{\theta})_{1}$,
$\lvf_{3}=\art(\eps_{r})_{2}$, $\lvf_{4}=\art(\eps_{\theta})_{2}$,
where $S_{\el}$ is the area of the $l$-th element and $(\eps_{r})_{\el},\;(\eps_{\theta})_{\el}$
are the uniform strain components in $\elem_{\el}$. The values of
the strain components within the elements are approximated as\begin{equation}
(\eps_{r})_{\el}=\frac{\vf_{\el+1}-\vf_{\el}}{r_{\el+1}-r_{\el}}\,,\quad(\eps_{\theta})_{\el}=\frac{\vf_{\el}+\vf_{\el+1}}{2m_{\el}}\,,\end{equation}
where $m_{l}=(r_{\el}+r_{\el+1})/2$ denotes the mean radius of the
$\el$-th element. The matrix $\amat$ is easily calculated to give\begin{equation}
\left[A\right]=\left[\begin{array}{ccc}
-\frac{\aro}{\rtwo-\ro}\vphantom{\int_{\frac{D}{D}a}^{\frac{D}{D}}} & \frac{\aro}{\rtwo-\ro} & 0\\
\vphantom{\int_{\frac{D}{D}a}^{\frac{D}{D}}}\frac{\aro}{2\mo} & \frac{\aro}{2\mo} & 0\\
\vphantom{\int_{\frac{D}{D}a}^{\frac{D}{D}}}0 & -\frac{\art}{\rth-\rtwo} & \frac{\art}{\rth-\rtwo}\\
\vphantom{\int_{\frac{D}{D}a}^{\frac{D}{D}}}0 & \frac{\art}{2\mt} & \frac{\art}{2\mt}\end{array}\right].\end{equation}
Instead of the optimal stress, it will be convenient to determine
\begin{equation}
\frac{1}{\optimum^{\opt}}=\min_{\vf}\frac{\sum_{\ldof}\abs{\amat_{\ldof\dof}\vf_{\dof}}}{\abs{\fvec_{\dof}\vf_{\dof}}}\,.\end{equation}
Since both numerator and denominator are homogeneous in the vector
$\vf$, we normalize it by conducting the search for the minimum over
all displacement vectors satisfying\begin{equation}
\fvec_{\dof}\vf_{\dof}=f_{3}\vf_{3}=1\end{equation}
so\begin{equation}
\vf_{3}=\frac{1}{2\pi\rth p}\,,\label{eq:normalizeW}\end{equation}
and we have to minimize\begin{equation}
\sum_{\ldof}\abs{\amat_{\ldof\dof}\vf_{\dof}}\end{equation}
over all values of $(\vf_{1},\vf_{2})$ subject to the condition (\ref{eq:normalizeW}).
Writing the sum above explicitly and using the fact that $\amat_{31}=\amat_{32}=0$,
we have\begin{equation}
\frac{1}{\optimum^{\opt}}=\min_{\vf_{2}}\braces{P+\abs{\amat_{3\dof}\vf_{\dof}}+\abs{\amat_{4\dof}\vf_{\dof}}},\label{eq:MinProb}\end{equation}
where\begin{equation}
P=\min_{\vf_{1}}\braces{\abs{\amat_{11}\vf_{1}+\amat_{12}\vf_{2}}+\abs{\amat_{21}\vf_{1}+\amat_{22}\vf_{2}}}.\end{equation}
As a function of $\vf_{1}$, the expression in the curly brackets
above is piecewise affine and attains its minimum at some point where
two adjacent line segments of its graph meet---at some value of $\vf_{1}$
where one of the absolute value terms vanishes. This gives \begin{equation}
P=\frac{\abs{\vf_{2}}\abs{\amat_{11}\amat_{22}-\amat_{12}\amat_{21}}}{\max\braces{\abs{\amat_{11}},\abs{\amat_{21}}}}=\frac{\abs{\vf_{2}}\abs{\amat_{11}\amat_{22}-\amat_{12}\amat_{21}}}{\abs{\amat_{11}}}\,,\end{equation}
as evidently $\abs{\amat_{11}}>\abs{\amat_{21}}$. Substituting the
values of the various components of the matrix we obtain\begin{equation}
P=2\pi(\rtwo-\ro)\abs{\vf_{2}}\end{equation}

We can turn back to the minimization of Eq.~(\ref{eq:MinProb}) where
now the minimum is attained at a value of $\vf_{2}$ where one of
the two absolute value terms vanishes or at the value where $P$ vanishes
(so $\vf_{2}=0$). Setting\begin{equation}
Q=P+\abs{\amat_{3\dof}\vf_{\dof}}+\abs{\amat_{4\dof}\vf_{\dof}},\end{equation}
for the case $\amat_{32}\vf_{2}+\amat_{33}\vf_{3}=0$, we have\begin{equation}
Q=Q_{1}=\frac{\abs{\vf_{3}}}{\abs{\amat_{32}}}\braces{\abs{\amat_{33}}\frac{D_{1}}{\max\braces{\abs{\amat_{11}},\abs{\amat_{21}}}}+D_{2}},\end{equation}
where $D_{1}=\amat_{11}\amat_{22}-\amat_{12}\amat_{21}$, and $D_{2}=\amat_{33}\amat_{42}-\amat_{43}\amat_{32}=2\pi\art$
are the determinants of the two submatrices. It follows that\begin{equation}
Q_{1}=2\pi(\rth-\ro)\abs{\vf_{3}}.\end{equation}
For the case $\vf_{2}=0$, we have\begin{equation}
Q=Q_{2}=(\abs{\amat_{33}}+\abs{\amat_{43}})\abs{\vf_{3}}=2\pi\rth\abs{\vf_{3}},\end{equation}
and for the case $\amat_{42}\vf_{2}+\amat_{43}\vf_{3}=0$, we have\begin{align}
Q=Q_{3} & =\abs{\vf_{3}}\braces{\frac{\abs{\amat_{43}}}{\abs{\amat_{42}}}\frac{D_{1}}{\max\braces{\abs{\amat_{11}},\abs{\amat_{21}}}}+\frac{D_{2}}{\abs{\amat_{42}}}}\nonumber \\
 & =2\pi(\rth+2\rtwo-\ro)\abs{\vf_{3}}.\end{align}
It follows that \begin{equation}
\frac{1}{\optimum^{\opt}}=\min_{\vf_{2}}\braces{Q_{1},Q_{2},Q_{3}}=Q_{1}\end{equation}
and substituting the normalization condition on $\vf_{3}$, we finally
obtain\[
\optimum^{\opt}=\frac{\rth}{\rth-\ro}\, p.\]
It is noted that this value corresponds to a uniform value of $\st_{\theta}$
that will balance the external loading for one half of the cylinder.

This result for a common engineering problem could have been possibly
obtained by direct analysis of the hollow cylinder. However, this
example was aimed at demonstrating the strength of the proposed procedure
and the ability to implement it in real life structures. Clearly,
aside from the practical significance of knowing the optimal stress
that may develop in a structure under given loading conditions, this
procedure can be used to assess the optimality of a given standard
design. Thus, comparing of the maximal stresses developing in a proposed
design with the optimal value obtained by application of the formulation
outlined above, one can estimate how much the design can be improved.

\section*{ACKNOWLEDGMENT}

This research was partially supported by The Paul Ivanier Center for
Robotics Research and Production Management and the Pearlstone Center
for Aeronautical Engineering Studies at Ben-Gurion University.

\appendix

\section{Appendix: Norm Preserving Extensions}


Let $\vs_{0}\subset\vs$ be a vector subspace and $\fc_{0}\from\vs_{0}\tto\reals$
a linear functional such that \begin{equation}
\fc_{0}(\vf_{0})\les\norm{\fc_{0}}\norm{\vf_{0}},\quad\text{for all}\quad\vf_{0}\in\vs_{0}.\end{equation}
Let $\vf_{1}$ be an element of $\vs-\vs_{0}$ and let $\vs_{1}$
be the vector space spanned by $\vs_{0}$ and $\vf_{1}$, i.e.,\[
\vs_{1}=\braces{\vf_{0}+a\vf_{1};\, a\in\reals,\,\vf_{0}\in\vs_{0}}.\]
We want to extend $\fc_{0}$ to a functional\[
\fc_{1}\from\vs_{1}\tto\reals,\]
such that\begin{align}
\fc_{1}(\vf_{0}) & =\fc_{0}(\vf_{0}),\quad\text{for all}\quad\vf_{0}\in\vs_{0}\\
\fc_{1}(\vf) & \les\norm{\fc_{0}}\norm{\vf},\quad\text{for all}\quad\vf\in\vs_{1}.\end{align}

Assuming $\fc_{1}$ satisfies the above requirements, then,\begin{align}
\fc_{1}(\vf_{0}+\vf_{1}) & =F_{1}(\vf_{0})+\fc_{1}(\vf_{1})\\
\norm{\fc_{0}}\norm{\vf_{0}+\vf_{1}} & \ges\fc_{0}(\vf_{0})+\fc_{1}(\vf_{1})\qquad\text{using $\norm{\fc_{1}}=\norm{\fc_{0}}.$}\end{align}
Thus,\begin{equation}
\fc_{1}(\vf_{1})\les\norm{\fc_{0}}\norm{\vf_{0}+\vf_{1}}-\fc_{0}(\vf_{0})\quad\text{for all}\quad\vf_{0}\in\vs_{0}.\label{eq:Cond1}\end{equation}
Similarly,\begin{align}
\fc_{1}(-\vf_{0}-\vf_{1}) & =-F_{1}(\vf_{0})-\fc_{1}(\vf_{1})\\
\norm{\fc_{0}}\norm{-\vf_{0}-\vf_{1}} & \ges-\fc_{0}(\vf_{0})-\fc_{1}(\vf_{1})\\
\norm{\fc_{0}}\norm{\vf_{0}+\vf_{1}} & \ges-\fc_{0}(\vf_{0})-\fc_{1}(\vf_{1}).\end{align}
Thus,\begin{equation}
\fc_{1}(\vf_{1})\ges-\norm{\fc_{0}}\norm{\vf_{0}+\vf_{1}}-\fc_{0}(\vf_{0})\quad\text{for all}\quad\vf_{0}\in\vs_{0}.\label{eq:Cond2}\end{equation}

We now show that the necessary conditions of Eqs. (\ref{eq:Cond1})
and (\ref{eq:Cond2}) are also sufficient. For any $\vf=\vf_{0}+a\vf_{1}\in\vs_{1}$
we have\begin{equation}
\fc_{1}(\vf_{0}+a\vf_{1})=a\fc_{0}(\vf_{0}/a)+a\fc_{1}(\vf_{1}),\label{eq:lin}\end{equation}
using $\fc_{1}(\vf_{0})=\fc_{0}(\vf_{0})$.

Now if $a>0$, we may use Eq. (\ref{eq:Cond1}) for $\vf_{0}/a$ in
the second term on the right of Eq.~(\ref{eq:lin}) to get\begin{align*}
\fc_{1}(\vf_{0}+a\vf_{1}) & \les a\fc_{0}(\vf_{0}/a)+a\paren{\norm{\fc_{0}}\norm{\vf_{0}/a+\vf_{1}}-\fc_{0}(\vf_{0}/a)}\\
 & =\norm{\fc_{0}}\norm{\vf_{0}+a\vf_{1}}.\end{align*}
Alternatively, for $a<0$, Equation (\ref{eq:Cond2}) may be rewritten
as\[
a\fc_{1}(\vf_{1})\les-a\norm{\fc_{0}}\norm{\vf_{0}+\vf_{1}}-a\fc_{0}(\vf_{0})\quad\text{for all}\quad\vf_{0}\in\vs_{0}.\]
When we substitute this for $\vf_{0}/a$ in Equation (\ref{eq:lin})
we obtain\begin{align*}
\fc_{1}(\vf_{0}+a\vf_{1}) & \les a\fc_{0}(\vf_{0}/a)-a\norm{\fc_{0}}\norm{\vf_{0}/a+\vf_{1}}-a\fc_{0}(\vf_{0}/a),\\
 & =-a\norm{\fc_{0}}\norm{\vf_{0}/a+\vf_{1}},\\
 & =\norm{\fc_{0}}\norm{\vf_{0}+a\vf_{1}},\end{align*}
where we used the fact that $(-a)>0$ in the last line above. This
completes the proof.

\end{document}